\documentclass[12pt]{article}

\usepackage{cite}
\usepackage{epsfig}
\usepackage{graphics}
\usepackage{inputenc}
\inputencoding{latin1}

\def\kt{$K_\perp~$}
\def\dcut{d_\mathit{cut}}
\def\ycut{y_\mathit{cut}}
\def\ecut{E_\mathit{cut}}
\def\ee{${e^+ e^-}~$}
\def\ktevent{\texttt{KtEvent}}
\def\ktlv{\texttt{KtLorentzVector}}

\begin{document}

\sloppy

\begin{titlepage}
{\par\raggedleft \texttt{MAN/HEP/2002/02}~\\
 \texttt{UCL/HEP 2002-02}\\
\texttt{October 2002}\par}
\bigskip{}

\bigskip{}
{\par\centering \textbf{\large KtJet : A C++ implementation of the 
{\boldmath \kt} clustering algorithm}\large \par}
\bigskip{}

{\par\centering J. M. Butterworth$^1$, J. P. Couchman$^1$, \\
 B. E. Cox$^2$ and B. M. Waugh$^1$\\
\par}
\bigskip{}

{\par\centering {\small \( ^{1} \)Department of Physics and Astronomy}\\
{\small University College London}\\
{\small Gower St. London WC1E 6BT}\\
{\small England}\small \par}

{\par\centering {\small \( ^{2} \)Department of Physics and Astronomy }\\
 {\small University of Manchester}\\
 {\small Manchester M13 9PL}\\
{\small England} \par}
\bigskip{}

\begin{abstract}
\noindent

A C++ implementation of the \kt jet algorithm for high energy particle
collisions is presented. The time performance of this implementation
is comparable to the widely used Fortran implementation.  Identical
algorithmic functionality is provided, with a clean and intuitive user
interface and additional recombination schemes.  A short description
of the algorithm and examples of its use are given.

\end{abstract}

{\it PACS:}  
\end{titlepage}

\section{Introduction}
\label{sec:intro}

This paper is intended to be an introduction to the use of KtJet. For a
detailed explanation of the $K_\perp$ clustering algorithm, its
properties, and the physics motivation behind it, the reader is
referred to \cite{Catani:1993hr} and references therein.  The KtJet
package implements all the features of the Fortran implementation of
the \kt algorithm \cite{Seymourfortran}. The philosophy throughout has
been to design an interface that users of the Fortran code will
recognise, but which exploits the advantages of object-oriented design.
Therefore the names and functions of the input parameters
are retained wherever possible. The KtJet library, examples of its use
and detailed documentation are all available on the KtJet website,
\texttt{http://www.ktjet.org/}, which also provides a link to a
CVS repository from which the latest version of the source code may be
obtained.

\section{The {\boldmath \kt} algorithm}

The \kt algorithm can function in several distinct ways depending on
the nature of the colliding beams and the physics to be studied. We
deal first with hadron-hadron and lepton-hadron collisions. There are two modes of
operation, the `inclusive' and `exclusive' modes. The difference
between the two cases, described in sections \ref{inclusive} and
\ref{exclusive}, is in the definition of the hard final state jets,
and the separation of these jets from the beam remnants.

For the analysis of \ee data, in contrast, there is of course no
concept of beam remnants. The algorithm proceeds in a very similar way
to that employed in a jet substructure analysis. We deal with these
two cases together in section \ref{ee}.

\subsection{The inclusive mode}
\label{inclusive}
The algorithm proceeds as follows: 

\begin{enumerate}

\item 
For every final state object\footnote{Final state objects could be,
for example, partons, hadrons, calorimeter cells, tracks etc.} $h_k$ and for
every pair $h_k$ and $h_l$, compute the resolution variables $d_{kB}$
and $d_{kl}$. The precise definition of these variables can be chosen
by the user (using the parameter \texttt{angle}), and will be
described in detail in section \ref{jrv}. They always have the
property that in the small angle limit they reduce respectively to the
squared relative transverse momentum of the object with respect to the
beam direction, and the squared relative transverse momentum of one
object with respect to the other. 

\begin{eqnarray}
\label{sa1}
  d_{kB} & \simeq & E_k^2 \theta_{kB}^2 \\
         & \simeq & k_{\perp kB}^2, {\rm for~} \theta \rightarrow 0 \nonumber \\
\label{sa2}
  d_{kl} & \simeq & {\rm min}(E_k^2,E_l^2)\theta_{kl}^2 \\
         & \simeq & k_{\perp kl}^2, {\rm for~} \theta_{kl} \rightarrow 0 \nonumber
\end{eqnarray}

At this stage, a dimensionless parameter $R$ is
introduced\cite{Ellis:tq}, which plays a radius-like role in defining
the extent of the jets. This is usually set to 1.0.

\item Scale the $d_{kB}$ by $R^2$  
\begin{equation}
 d_k = d_{kB} R^2
\end{equation} 

\item Find the smallest value among the $d_k$ and $d_{kl}$. If a
$d_{kl}$ is the smallest, $h_k$ and $h_l$ are combined into a single
object with momentum $p_{(kl)}$ according to a user specified
recombination scheme (parameter \texttt{recom}) which will be
described in section \ref{rs}. As an example, \texttt{recom = 1} would
correspond to 4-vector addition. If a $d_k$ is the smallest, object
$k$ is defined to be a jet and is removed from the list of objects
to be merged.

\item Repeat until all objects have been included in jets.

\end{enumerate}

\subsection{The exclusive mode}
\label{exclusive}

In this mode, the algorithm separates the `hard final state' from the
soft `beam remnants' explicitly. Jets are defined in the hard final
state by a stopping parameter $\dcut$ with dimensions of energy squared. 
\begin{enumerate}

\item 
$d_{kB}$ and $d_{kl}$ are defined as in section \ref{inclusive}. 

\item 
Find $d_{min}$, the smallest value among the $d_{kB}$ and $d_{kl}$. If
$d_{min} > \dcut$, all remaining objects in the event will be
classified as jets, and the algorithm is complete.

\item 
If a $d_{kl}$ is the smallest, $h_k$ and $h_l$ are combined into a
single object with momentum $p_{(kl)}$ according to a user specified
recombination scheme (parameter \texttt{recom}). If a $d_{kB}$ is the
smallest, object $k$ is included in a `beam jet' and removed from the list.
Go back to (2).

\end{enumerate}
Note that the stopping
parameter $\dcut$ defines the hard scale of the process
($\Lambda_{QCD}^2 \ll \dcut \le s$, where $\sqrt s$ is the
centre-of-mass energy). One can therefore think of the hard subprocess
being factorised from the low-$p_\perp$ scattering fragments, which
are thrown away into the beam jets.

Instead of specifying the stopping scale $\dcut$, one can choose to
stop merging when a given number of jets is reached.

\subsection{Subjet analysis and {\boldmath \ee} mode}
\label{ee}

Once the hadronic final state has been decomposed into jets, the
structure of the jets themselves can be investigated. This is commonly
known as a subjet analysis. The procedure is physically (and
practically) identical to that employed in the analysis of an \ee
event. In the \ee case, all final state objects are used as input to
the algorithm. In a subjet analysis on a particular jet, only the
particles within that jet are used as input. The steps are as follows;

\begin{enumerate}
\item Define a resolution parameter
\begin{equation}
\label{ycutdef}
 \ycut = Q_0^2 / E_\mathit{cut}^2. 
\end{equation}

The parameter $\ecut^2$ may be chosen by the user, but is
conventionally taken to be the square of the total energy of the \ee event
(or the $p_t$ of the jet) in the frame in which the algorithm is run. These are the
default settings in KtJet.

\item For each pair of objects $h_k$ and $h_l$, construct the rescaled
resolution variable $y_{kl}$
\begin{equation}
\label{ycut}
y_{kl}=d_{kl}/E_\mathit{cut}^2
\end{equation}

where $d_{kl}$ is defined as in section \ref{inclusive}. 

\item Find $y_{min}$, the smallest of the $y_{kl}$. If $y_\mathit{min} < \ycut$, $h_k$
and $h_l$ are combined into a single object with momentum $p_{(kl)}$
according to a user specified recombination scheme (parameter
\texttt{recom}).
\item Repeat the procedure until all pairs of objects have $y_{kl} >
\ycut$. The remaining objects are called subjets.

\end{enumerate}

As described in section~\ref{exclusive}, one can choose to stop merging when
a given number of jets is reached, instead of specifying the stopping scale $\ycut$.

\subsection{Jet resolution variables}
\label{jrv}

The small-angle behaviour of the resolution variables $d_{kB}$ and
$d_{kl}$ given in equations \ref{sa1} and \ref {sa2} controls the
behaviour of the \kt algorithm in the soft and collinear limits. Away
from the collinear limit, the user has a choice of angular
definitions, controlled by the parameter \texttt{angle}. Note that the
resolution variables are defined such that they remain monotonic
functions of angle at large angles.  Here, and throughout this paper,
we use $\eta$ to denote true rapidity, i.e.\ $\eta =
\frac{1}{2}\ln\frac{E+p_z}{E-p_z}$.

\subsubsection{The angular scheme, \texttt{angle = 1}}
\begin{eqnarray}
d_{kB} & = & 2E^2_k(1-cos\theta_{kB}), \nonumber \\ 
d_{kl} & = & 2 \min(E^2_k,E^2_l)(1-cos\theta_{kl}).
\end{eqnarray} 
This definition was originally proposed for jets in DIS \cite{Catani:1992zp} 
(where the \kt algorithm should be implemented in the Breit frame). 

\subsubsection{The $\Delta R$ scheme, \texttt{angle = 2}}
\begin{eqnarray}
d_{kB} & = & p_{tk}^2, \nonumber \\ 
d_{kl} & = & \min(p^2_{tk},p^2_{tl})R^2_{kl},
\end{eqnarray} 
where 
\begin{equation}
R^2_{kl}=(\eta_k-\eta_l)^2+(\phi_k-\phi_l)^2.
\end{equation}
This definition corresponds to that used in cone algorithms, and is
the most common choice for hadron-hadron collisions.

\subsubsection{The QCD emission scheme, \texttt{angle = 3}}

An alternative definition of $R^2_{kl}$, motivated by the form of the
QCD matrix elements for multi-parton emissions:
\begin{equation}
R^2_{kl}=2[cosh(\eta_k-\eta_l)-cos(\phi_k-\phi_l)].
\end{equation}

\subsection{Recombination schemes}
\label{rs}

The recombination scheme defines how two objects $h_k$ and $h_l$ are
merged into a single object with 4-momentum $p_{(kl)}$. Five schemes
are available, selected by the parameter \texttt{recom}.

\subsubsection{The {\boldmath $E$} scheme, \texttt{recom=1}}
Simple 4-vector addition:
\begin{equation}
p_{kl}=p_k+p_l.
\end{equation}
This procedure results in massive final state jets.

\subsubsection{The {\boldmath $p_t$} scheme, \texttt{recom=2}}
\label{ptscheme}
\begin{eqnarray}
p_{t(kl)} & = & p_{tk}+p_{tl}, \nonumber \\
\eta_{kl} & = & \frac{p_{tk}\eta_k+p_{tl}\eta_l}{p_{t(kl)}}, \nonumber \\
\phi_{kl} & = & \frac{p_{tk}\phi_k+p_{tl}\phi_l}{p_{t(kl)}}.
\end{eqnarray} 

This definition constrains only the 3 spatial components of
the object's 4-vector.
The combined object is made massless by setting its energy equal to the magnitude
of its 3-momentum.
If massive objects are input (for instance particle 4-vectors from a simulated event)
they are made massless in the same way before the $d_{kB}$ and $d_{kl}$ are calculated.
Compare the $E_t$~scheme (section~\ref{etscheme}).

\subsubsection{The {\boldmath $p_t^2$} scheme, \texttt{recom=3}}
\begin{eqnarray}
p_{t(kl)} & = & p_{tk}+p_{tl}, \nonumber \\
\eta_{kl} & = & \frac{p^2_{tk}\eta_k+p^2_{tl}\eta_l}{p^2_{tk}+p^2_{tl}}, \nonumber \\
\phi_{kl} & = & \frac{p^2_{tk}\phi_k+p^2_{tl}\phi_l}{p^2_{tk}+p^2_{tl}}.
\end{eqnarray} 

This definition constrains only the 3 spatial components of the
object's 4-vector.
The energy is made equal to to the magnitude
of its 3-momentum, thus making the combined object massless.
Note also that this definition of the $p_t^2$ scheme is that used in the Fortran
implementation of the algorithm \cite{Seymourfortran}. It is not equivalent to the 
monotonic $p_t^2$ scheme defined in equations 16 and 17 of reference \cite{Catani:1993hr}. 
We discuss the issue of monotonicity further in section \ref{monot}. 

\subsubsection{The {\boldmath $E_t$} scheme, \texttt{recom=4}}
\label{etscheme}
\begin{eqnarray}
E_{t(kl)} & = & E_{tk}+E_{tl}, \nonumber \\
\eta_{kl} & = & \frac{E_{tk}\eta_k+E_{tl}\eta_l}{E_{t(kl)}}, \nonumber \\
\phi_{kl} & = & \frac{E_{tk}\phi_k+E_{tl}\phi_l}{E_{t(kl)}}.
\end{eqnarray} 

For massless input objects this definition is identical to the $p_t$
scheme (section \ref{ptscheme}). It differs solely in the way it deals
with massive input objects. If massive objects are input, the $p_t$ scheme uses
their transverse momentum, whereas the $E_t$ scheme uses the
transverse energy $E\sin\theta$. This can have a significant effect
for low $E_t$ jets, particularly for steeply falling distributions
close to a cut-off value. All combined objects are massless in both
cases.

\subsubsection{The {\boldmath $E_t^2$} scheme, \texttt{recom=5}}
\begin{eqnarray}
E_{t(kl)} & = & E_{tk}+E_{tl}, \nonumber \\
\eta_{kl} & = & \frac{E^2_{tk}\eta_k+E^2_{tl}\eta_l}{E^2_{tk}+E^2_{tl}}, \nonumber \\
\phi_{kl} & = & \frac{E^2_{tk}\phi_k+E^2_{tl}\phi_l}{E^2_{tk}+E^2_{tl}}.
\end{eqnarray} 

This bears the same relationship to the $p_t^2$ scheme that the $E_t$
scheme bears to the $p_t$.  All combined objects are massless.

\section{Practical implementation of the algorithm}
\label{practical}
KtJet uses the \texttt{HepLorentzVector} class of the CLHEP package
\cite{clhep}. The input 4-vectors (the objects on which the algorithm
will be run) may be \texttt{HepLorentzVector}s or \ktlv s.  
The \ktlv~class, which inherits from \texttt{HepLorentzVector}, carries
an internal index which allows the user to determine, for example, to
which final state jet a particular input particle belongs.
Examples of the use of \texttt{KtLorentzVector} are given in section~\ref{ktlvmethods}.
The output 4-vectors (the jets) are instantiated as \ktlv s.
The \ktlv~class has the constructors
\begin{verbatim}
KtLorentzVector(const HepLorentzVector &);
KtLorentzVector(const KtLorentzVector &);
KtLorentzVector(float Px, float Py, float Pz, float E);
\end{verbatim}

The first step in running the algorithm is the creation of a
\texttt{KtEvent} object. There are separate constructors to run the
algorithm in inclusive and exclusive modes respectively: \\
\\\texttt{KtEvent(const std::vector<KtLorentzVector> particles \&,
  int type, int angle, int recom, float rparameter);} \\
\\\texttt{KtEvent(const std::vector<KtLorentzVector> particles \&,
  int type, int angle, int recom);}\\ \\
There are also versions of these constructors taking vectors of
\texttt{HepLorentzVector}s instead of \texttt{KtLorentzVector}s.
The parameter \texttt{type} should be set to the colliding beam type,
as defined in table 1. The parameters \texttt{angle} and
\texttt{recom} are described in sections \ref{jrv} and
\ref{rs}. \texttt{rparameter} is the parameter $R$ defined in section~\ref{inclusive}
and should almost always be set to $1.0$,
although values smaller than $1.0$ have been suggested for certain
applications \cite{Seymour:1993mx,Seymour:2000yy}.
\begin{table}
\label{table1}
\begin{center}
\begin{tabular}{|l|l|l|} \hline
\texttt{type}  & Beam & Comments \\ \hline
 1 & $ee$ &  \\ \hline
 2 & $ep$ & $p$ in $-z$ direction \\ \hline
 3 & $pe$ & $p$ in $+z$ direction  \\ \hline
 4 & $pp$ &  \\ \hline
\end{tabular}
\end{center}
\caption{Possible input values for \texttt{type}.}
\end{table}
\subsection{KtEvent methods}
\label{ktmeth}
For the inclusive case, the final state jets are now fully defined,
and the jets can be recovered using the methods described below. In
the exclusive case, the final state jets are defined by setting either
the stopping parameter $\dcut$: \\
\\\texttt{void findJetsD(float dcut);}\\ \\
or by forcing the final state to decompose into $N$ jets: \\
\\\texttt{void findJetsN(int N);}\\ \\
The $d_{min}$ value at which the final state changes from $N+1$ to  
$N$ jets can be recovered using the method \\
\\\texttt{float getDMerge(int N) const;} \\

In the case of an \ee analysis, the variable $\ycut$ may be
used. The final state jets are defined by the method \\ \\
\texttt{void {findJetsY(float yCut);}} \\ \\
and the $y_\mathit{min}$ value at which the final state changes from
$N+1$ to $N$ jets can be recovered using the method \\ \\
\texttt{float getYMerge(int N) const;} \\ \\
By default, $\ycut$ is defined as in equation \ref{ycutdef}, where $\ecut$ is
taken to be the total energy in the event. The $\ecut$ value can be
set by the user if required: \\ \\
\texttt{void {setECut(float ECut);}} \\

The \texttt{KtEvent} object has the following methods for recovering its final state jets:\\ 
\\\texttt{std::vector<KtLorentzVector> getJets() const;} \\
Returns jets without sorting. \\
\\\texttt{std::vector<KtLorentzVector> getJetsE() const;} \\
Returns jets in order of decreasing energy. \\
\\\texttt{std::vector<KtLorentzVector> getJetsEt() const;} \\
Returns jets in order of decreasing transverse energy. \\
\\\texttt{std::vector<KtLorentzVector> getJetsPt() const;} \\
Returns jets in order of decreasing transverse momentum.\\
\\\texttt{std::vector<KtLorentzVector> getJetsRapidity() const;} \\
Returns jets in order of decreasing rapidity.\\
\\\texttt{std::vector<KtLorentzVector> getJetsEta() const;} \\
Returns jets in order of decreasing pseudorapidity.\\
\\\texttt{KtLorentzVector getJet(const KtLorentzVector \&) const;} \\
Returns the final state jet which contained the input particle
\ktlv. Note that this method is only available if
\texttt{KtLorentzVector}s were used as input in the \texttt{KtEvent}
constructor.\\
\\\texttt{std::vector<const KtLorentzVector*> \& getConstituents() const;} \\
Returns a vector of pointers to all the input objects in the \texttt{KtEvent}. \\
\\\texttt{std::vector<KtLorentzVector> copyConstituents() const;} \\
Returns a vector of copies of the input objects in the \texttt{KtEvent}. \\
\\\texttt{int getNConstituents() const;} \\
Returns the number of objects in the \texttt{KtEvent}.\\
\\The parameters of a particular \ktevent~can be retrieved using the methods\\
\texttt{float getETot() const;} \\
\texttt{int getType() const;}\\
\texttt{int getAngle() const;} \\
\texttt{int getRecom() const;} \\
\texttt{float getECut() const;} \\
\texttt{bool isInclusive() const;}\\

\subsubsection{Monotonicity}
\label{monot}        
Although in practice most users will not have to face the issue, it is
worth bearing in mind that recombination schemes 1 to 5 are not
guaranteed to lead to monotonic resolution variables, i.e.\ it is NOT
necessarily true that  ${\rm min}\{d_{kB},d_{kl}\} \le {\rm
min}\{d^{\prime}_{kB},d^{\prime}_{kl}\}$, where $d_k$ and $d_k^{\prime}$ are the
resolution variables before and after recombination respectively. This
means that, physically, the question ``How many jets are there at a
particular scale $d$?'' may not have a unique answer.  In KtJet, as
should be clear from section \ref{exclusive}, for a particular value
of $\dcut$ set by \texttt{findJetsD}, the largest value of $N$ (i.e.\
the first place at which $d_\mathit{min} > \dcut$) will be returned by
\texttt{getNJets}.  Similarly, there may be no value of $\dcut$ for
which a particular event has $N$ jets. There will however always be a
$d_{min}$ value at which the event changed from $N+1$ to $N$ jets,
which is the value returned by \texttt{getDMerge}.  The above
discussion also applies to the $y$ variables and associated methods.

\subsection{KtLorentzVector Methods}
\label{ktlvmethods}
These methods would normally be used to investigate the structure and
constituents of final state jets defined in a particular \ktevent.\\ 
\\\texttt{const std::vector<const KtLorentzVector*> \& getConstituents() const;} \\
Returns a reference to the vector of pointers to all the objects in the \ktlv. \\
\\\texttt{std::vector<KtLorentzVector> copyConstituents() const;} \\
Returns a vector of copies of the objects in the \ktlv. \\
\\\texttt{int getNConstituents() const;} \\
Returns the number of objects in the \ktlv.\\
\\\texttt{bool contains (const KtLorentzVector \&) const;}\\
Check if a jet contains a particular object. For example,
\\\texttt{if (JET.contains(PARTICLE)) \{ /* do something */ \}}\\
where \texttt{JET} is the \ktlv~of a final state jet and
\texttt{PARTICLE} is the \ktlv~of an input object in the \ktevent. \\
\\\texttt{KtLorentzVector \& operator+= (const KtLorentzVector \&);} \\
Adds a \texttt{KtLorentzVector} constituent to a jet using the $E$~scheme (4-vector addition)
and maintains an internal record of constituents (so for example the
\texttt{getConstituents} method will work on the resulting \ktlv).\\
\\\texttt{void add (const KtLorentzVector \&, int recom);} \\
Adds a \texttt{KtLorentzVector} constituent to a jet using any of the available recombination schemes
and maintains an internal record of constituents.

Note that the momentum of a jet (\texttt{KtLorentzVector}) will be given according to the
recombination scheme used in its construction. If the user wishes to reconstruct the
momenta according to a different recombination scheme (for example, to
recover the mass for jets which were found using a massless recombination scheme) the
\texttt{getConstituents()} or \texttt{copyConstituents()} methods may be used.  
The constituents can then be recombined in a new scheme using the \texttt{add} method.

\section{Subjet analysis}

A subjet analysis is performed on a particular final state jet by
constructing a new \ktevent~object: \\ \\
\texttt{KtEvent(const KtLorentzVector jet \&, int type, int angle, int recom);} \\ \\
The subsequent analysis and methods are identical to those described above
for the $e^+ e^-$ case: for example, subjets can be defined using the
\texttt{findJetsY} or \texttt{findJetsN} methods.  $\ecut$ is taken
to be the transverse momentum of the jet, although this can be changed by the
user using the \texttt{setECut} method.

\section{Advanced features of the KtLorentzVector class}

The \ktlv~class has several features which are used internally by
KtJet, but which may be useful for the user. 

A \ktlv~can be constructed in two different ways. When instantiated
using the constructor \texttt{KtLorentzVector(const HepLorentzVector
\&)} (or \texttt{KtLorentzVector(float Px,float Py,float Pz,float
E)}), the \ktlv~is simply a copy of a single \texttt{HepLorentzVector}
with an internal index added, i.e.\ it has no constituents. This index
is then used, for example, in the \texttt{getJet} method, to
ascertain to which final state jet a particular input
\ktlv~belongs. There are two methods which compare the {\it index} of
two \texttt{KtLorentzVector}s:
\begin{verbatim}
bool operator== (const KtLorentzVector &) const;
bool operator!= (const KtLorentzVector &) const;
\end{verbatim}
These would be useful if, for example, one wanted to ascertain whether
the highest $p_t$ particle in a particular jet was a given input \ktlv. 
A \ktlv~may also be instantiated using the constructor \\ 
\\\texttt{KtLorentzVector();}\\ \\
The \ktlv~can then be built up by adding other \ktlv s to it using
either the assignment operator \texttt{+=} or the \texttt{add} method
described in section \ref{ktlvmethods}. The \ktlv~will then also carry
a list of pointers to the constituent \ktlv s. To find out whether a
\ktlv~has constituents, use the method\\
\\\texttt{bool isJet() const;} \\ \\
which will be true for a \ktlv~with a constituent list.\\

\section{Using KtJet}

In this section we demonstrate a simple example analysis using the KtJet
library. We run the inclusive \kt algorithm on a proton-proton
collision and do a subjet analysis on the highest $E_t$ jet in the
event.

\begin{verbatim}
// This header file must be included.
#include "KtJet.h"

std::vector<KtLorentzVector> jetvec;

// Loop over input particles.
for (int i=0; i<npart; i++) {
  KtLorentzVector  r = KtLorentzVector(p[i][0],p[i][1],p[i][2],p[i][3]);
  jetvec.push_back(r);
}
// We now have a vector of KtLorentzVector, jetvec,
// containing all the particle 4-vectors, p, in the event.

// Run the inclusive KT algorithm in PP mode using the covariant E-scheme
// (type=4, angle=2, recom=1, rparameter=1).
KtEvent ev(jetvec,4,2,1,1.0);

// Get jets, sorted in Et.
std::vector<KtLorentzVector> jetsEt = ev.getJetsEt();

// Perform subject analysis on highest Et jet (angle=2, recom=1).
KtEvent jet1(jetsEt[0],2,1);

// Decompose jet1 into 2 subjets.
jet1.findJetsN(2);

// Get the KtLorentzVectors of the 2 subjets
std::vector<KtLorentzVector> subjetsEt = jet1.getJetsEt();

// Write out the pseudorapidity of the highest Et subjet.
cout << subjetsEt[0].eta() << endl;
\end{verbatim}

\section{Optimisation issues}

The time taken to process events grows with the multiplicity $n$
of the input objects as $n^3$. As the multiplicity grows with energy, and
current and future colliders have ever higher event rates, timing
issues are therefore of increasing practical importance.

A few compilation options are worth considering if speed is critical.
A compile option to switch from single to double precision arithmetic
is available, but single precision is used by default as
this increases the speed by around 10\%. In gcc, the
option \texttt{-O2} is essential; higher or lower optimisation levels
seriously degrade performance. Choice of architecture is also
important. For example using the gcc compiler flag
\texttt{-march=athlon} on an Athlon processor gives a performance gain
of around 5\%. More details on performance issues are given at
\texttt{http://www.ktjet.org/}.

\section{Adding new functionality}

KtJet interfaces with its recombination and jet resolution schemes via purely
abstract base classes called \texttt{KtRecom} and \texttt{KtDistance} respectively. The
packaged code comes supplied with 5 Recombination and 3 jet resolution schemes,
shown in table 2 along with their \texttt{angle} and \texttt{recom} flags.

Users can define their own recombination and jet resolution scheme
classes which inherit from the base classes \texttt{KtRecom} and
\texttt{KtDistance} and use these instead of the supplied schemes.

To use their own scheme the user needs to pass pointers to the base classes,
instantiated as their own scheme objects, in the \texttt{KtEvent} constructor
instead of the integer flags. 
\begin{table}
\label{table2}
\begin{center}
\begin{tabular}{|l|l|l|} \hline
\texttt{Flag}  & Class Name  \\ \hline
\texttt{angle = 1} & \texttt{KtDistanceAngle(int type)}   \\ \hline
\texttt{angle = 2} & \texttt{KtDistanceDeltaR(int type)} \\ \hline
\texttt{angle = 3} & \texttt{KtDistanceQCD(int type)}  \\ \hline
\texttt{recom = 1} & \texttt{KtRecomE}   \\ \hline
\texttt{recom = 2} & \texttt{KtRecomPt}   \\ \hline
\texttt{recom = 3} & \texttt{KtRecomPt2}   \\ \hline
\texttt{recom = 4} & \texttt{KtRecomEt} \\ \hline
\texttt{recom = 5} & \texttt{KtRecomEt2} \\ \hline
\end{tabular}
\end{center}
\caption{The names of the jet resolution and recombination classes included in KtJet.}
\end{table}

As an example, if the user defines a new jet resolution scheme in the
class \texttt{KtDistanceNew}, and wishes to use the $E$ recombination
scheme in an $ep$ collision, the inclusive \texttt{KtEvent} constructor
should be called as follows:
\begin{verbatim}
KtDistance* distance_scheme = new KtDistanceNew(2);
KtEvent ev(particles, 2, distance_scheme, 1, rparameter);
\end{verbatim}

If the user defines a new recombination scheme in the class
\texttt{KtRecomNew}, and wishes to use the \texttt{angle=2} scheme in
a $pp$ collision, the inclusive \texttt{KtEvent} constructor should be called
as follows:
\begin{verbatim}
KtRecom* recom_scheme = new KtRecomNew();
KtEvent ev(particles, 4, 2, recom_scheme, rparameter);
\end{verbatim}

Users are invited to submit any such additions to the authors for
inclusion in future releases.

\section*{Acknowledgements}
We would like to thank Mike Seymour for useful discussions, suggestions
and encouragement throughout the project.
This work was funded in the UK by PPARC and HEFCE.

\end{document}